\documentclass[review]{elsarticle}

\usepackage{dcolumn,braket}
\usepackage[version=3]{mhchem}
\usepackage{graphicx}

\journal{Journal of Luminescence}
\newcommand\cco{\ce{Cu2O} }

\begin{document}

\begin{frontmatter}

\title{Evaluation of defects in cuprous oxide through exciton luminescence imaging}

\author[phys]{Laszlo Frazer\corref{mycorrespondingauthor}}
\cortext[mycorrespondingauthor]{Corresponding author}
\ead{jl@laszlofrazer.com}

\author[phys]{Erik J. Lenferink}
\author[chem]{Kelvin B. Chang}
\author[chem,anl]{Kenneth R. Poeppelmeier}
\author[phys]{Nathaniel P. Stern}
\author[phys,eecs]{John B. Ketterson}

\address[phys]{Department of Physics, Northwestern University, 2145 Sheridan Road, Evanston, IL 60208, USA}
\address[chem]{Department of Chemistry, Northwestern University, 2145 Sheridan Road, Evanston, IL 60208, USA}
\address[anl]{Chemical Sciences and Engineering Division, Argonne National Laboratory, 9700 South Cass Avenue, Argonne, IL 60439, USA}
\address[eecs]{Department of Electrical Engineering and Computer Science, Northwestern University, 2145 Sheridan Road, Evanston, IL 60208, USA}

\begin{abstract}
	The various decay mechanisms of excitons in cuprous oxide (\ce{Cu2O}) are highly sensitive to defects which can relax selection rules.  Here we report cryogenic hyperspectral imaging of exciton luminescence from cuprous oxide crystals grown via the floating zone method showing the samples have few defects.  Some locations, however, show strain splitting of the \emph{1s} orthoexciton triplet polariton luminescence.  Strain is reduced by annealing.  In addition, annealing causes annihilation of oxygen and copper vacancies, which leads to a negative correlation between luminescence of unlike vacancies.
\end{abstract}


\end{frontmatter}

\section{Introduction}

Owing to the positive parity of the valence and conduction bands, the primary luminescence mechanism in cuprous oxide is emission of two negative parity particles.  The strongest process is the decay of an orthoexciton into a 0.014 eV $^2\Gamma_{12}^-$ phonon and a photon \cite{petroff1975study}.  In addition, there is polariton luminescence from the quadrupole quantum mixing of the orthoexciton with a photon \cite{frohlich1991coherent}.  Under mechanical stress, the orthoexciton triplet state is split, and so is its luminescence \cite{sandfort2011paraexcitons}.  The paraexciton can also begin to produce direct luminescence under stress owing to breaking of the positive parity of the band structure \cite{sandfort2011paraexcitons,lin1993bose}.  These properties are unique to \ce{Cu2O} and \ce{Ag2O}, the two cuprite structure materials.

On the one hand, deliberate application of strain to cuprous oxide can be useful for increasing luminescence \cite{trauernicht1986thermodynamics}, trapping excitons in a potential \cite{yoshioka2011transition,trauernicht1986thermodynamics}, initiating transverse (negative parity) phonon emission \cite{yoshioka2013generation}, or distinguishing otherwise degenerate states \cite{sun2002strain}.  On the other hand, unwanted strain increases Auger recombination \cite{denev2002stress,wolfe2005new,laszlo2013unexpectedly} and breaks the unique symmetry of cuprite structured crystals.  Under practical experimental conditions, cuprous oxide is metastable.  The thermodynamic phase is cupric oxide \cite{schmidt1974growth} which appears in inclusions when cuprous oxide crystals are cooled too quickly \cite{chang2013removal,koohpayeh2008optical,dabkowska2010crystal,ludge2010floating}.  In this report, interfacial stress between the cuprous oxide and the cupric oxide inclusions is a source of strain \cite{haydar1980excitonic}.

Luminescence at 1.9477 eV has previously been described as ``very weak'' and resolvable only at temperatures below 4.2 K \cite{petroff1975study}.  Very little is known about this luminescence, except that it is thought to be extrinsic.  This is not discussed in most luminescence studies, probably because most CCD spectrometers have narrow spectral ranges that cannot cover this line and the better known luminescence lines simultaneously.

Cuprous oxide is nonstoichiometric \cite{raebiger2007origins,isseroff2013electronic,zouaghi1972near}.  Excitons bind to vacancies in the crystal and then undergo radiative decay \cite{koirala2013correlated}.  The dominant luminescence is from copper vacancies $V_{\ce{Cu}}^-$.  There are also two types of oxygen vacancy luminescence, from $V_{\ce{O}}^{1+}$ and $V_{\ce{O}}^{2+}$.  Vacancies are useful for increasing the conductivity of cuprous oxide \cite{o1961electrical,toth1961electrical,ochin1984thermodynamic,porat1995defect}.  However, the luminescence produced by vacancies indicates that vacancies reduce the lifetime of excitons \cite{koirala2013correlated,li2013engineering}.  The simultaneous existence of both copper and oxygen vacancies occurs because real samples are not perfectly equilibriated.  Ultimately, low temperature equilibration conditions should lead to elimination of the minority defect.

In this paper, we investigate the spatial distribution of defects in large cuprous oxide crystals developed for optical applications \cite{frazer2014third} to demonstrate the high quality of floating zone synthesis methods.  We also describe the underlying defect/exciton and defect/defect interactions.

\section{Materials and Methods}

\subsection{Samples}
A crystal of cuprous oxide was grown using the floating zone method as previously described \cite{chang2013removal}.  The starting material was 99.9\% \ce{Cu} rods with a 5 mm diameter.  The crystal was grown in air at 7 mm/h using two oxidized rods rotating at 7 rpm in opposite directions.  In this study, an as-grown sample and a sample annealed at 1045 $^\circ$C for 5 days with a 5 $^\circ$C/minute ramp rate are compared.  The samples were polished.  In Reference \cite{chang2013removal}, these samples are shown as Fig. 8 (a, d) and contribute to Fig. 7.

\subsection{Luminescence Measurements}

The samples were placed in an optical microscopy cryostat (Montana Instruments Cryostation).  Thermal contact was established with VGE-7031 varnish.  The stage temperature was about 5.7 K, with a stability of about 5 mK.  The temperature dependence of cuprous oxide luminescence has been well characterized \cite{ito1997detailed}.  The samples were in vacuum.  A scanning microscope with a 50X, 13 mm working distance objective was used to image the samples through a single window.  The samples were illuminated through the objective using a reflection from a beamsplitter with 4.3 mW (measured between the objective and the beamsplitter) of 532 nm light from a Coherent Verdi G18 laser.  Luminescence from the sample passed back through the objective, was partially transmitted through the beamsplitter, passed through a 532 nm long-wavelength-passing dielectric filter, and was collected in a fiber.  The spectrum was recorded with an Andor 303 mm focal length Czerny-Turner spectrograph and DU420A-BEX2-DD CCD camera.

The background was subtracted from the spectrum.  During background measurements, the laser beam was blocked.  To obtain a consistent spectrum, it was necessary to wait approximately 15 minutes for the stage temperature to stabilize after the sample was illuminated.  

Exciton luminescence for each sample was recorded over a 25$\times$25 square grid of locations with a spacing of 20 $\mu$m for 30 seconds using a 1200 grooves/mm grating.  This process took about 7 hours per sample.  For the exciton luminescence, the manufacturer specifies an instrument resolution of 372 $\mu$eV.

Vacancy luminescence, which has no spectrally narrow features, was recorded using a 150 grooves/mm grating over a square grid of locations with a spacing of 28 $\mu$m.  Since a more efficient, coarser grating was used, only one second was required to collect a good vacancy luminescence spectrum and it was possible to easily sample a larger number of locations.

Our previous study \cite{chang2013removal} was designed to compare room temperature copper vacancy luminescence across samples.  To achieve this, each sample was placed at the same distance along the optical axis from the objective.  In this experiment, we investigate variations in luminescence within samples, including lines best observed at temperatures below 10 K  \cite{ito1997detailed}.  The design of the cryostation does not permit each sample to be placed at precisely the same location along the optical axis, so direct comparisons of luminescence brightness between samples are not possible.  When changing between samples, the microscope was refocused to optimize the efficiency with which the luminescence was collected.  All measurements reflect conditions near the sample surface because the laser light has a short absorption length \cite{baumeister1961optical}.  Since excitons and exciton polaritons propagate differently \cite{frohlich1991coherent}, the different types of luminescence come from slightly different volumes, with the highest density and greatest brightness occurring at the laser spot. 

\subsection{Analysis Methods}
In summary, the phonon-assisted luminescence is modeled.  The residuals are used to determine the brightness, energy, and width of the orthoexciton polariton luminescence.  Analysis is performed for each location in the hyperspectral image.

Excitons in these conditions are Maxwell-Boltzmann distributed \cite{o2000relaxation,shen1997dynamics,snoke1990picosecond}.  They can decay into a phonon/photon pair if the phonon has negative parity.  There are also three and four particle complications \cite{petroff1975study}.  Phonon-assisted luminescence was modeled using the Maxwell-Boltzmann equation for the spectral irradiance $I$ as a function of energy $E$ \cite{gross1966free}:  
\begin{align}
	I(E)&=A\left(\left|E-E_c\right|\right)^{\frac12}e^{-\frac{E-E_c}{k_BT}};\label{mb}
\end{align}
Here $A$ is the brightness, $E_c$ is the orthoexciton ground state energy minus the phonon energy, and $k_B$ is the Boltzmann constant.  We only analyzed the most efficient luminescence, which is linked to emission of the $^2\Gamma_{12}^-$ phonon.  The other phonon-assisted luminescence peaks do not significantly overlap with this peak at 5.7 K \cite{ito1997detailed}.  We also tried adding a constant background term to the model, but concluded that the background was adequately accounted for by the experimental background subtraction.  The remaining error in the spectrum baseline after the background subtraction procedure contributes to error in the results.  For each of 625 locations on each sample, the model was applied between 2.0180 and 2.0300 eV to avoid the region $E<E_c$, where the model is invalid, and to avoid the 2.0318 eV orthoexciton polariton luminescence which could skew the fit.

The orthoexciton polariton luminescence, which intrinsically has a very narrow spectral width \cite{schmutzler2013signatures,frohlich1991coherent}, can typically be modeled by a Gaussian.  The width of the Gaussian is determined by the spectrometer resolution.  In this study, we are searching for deviations from the typical luminescence spectrum owing to sample defects.  These deviations can cause a Gaussian model, or any other single peaked model, to fail, so for each location we computed the residual brightness after subtracting Eq. \ref{mb}. Next we computed the sum over energies of the residual brightness, the residual brightness square weighted mean energy, and the likewise weighted standard deviation energy of the spectrum between 2.0300 and 2.0335 eV.  These three statistics describe the brightness, energy, and width of the orthoexciton polariton luminescence respectively.  No assumption is made about the structure of the polariton luminescence in the analysis.

For the defect linked luminescence near 1.95 eV we do not have a line shape model.  To evaluate it, we summed brightness in the spectrum between 1.9443 and 1.9478 eV.

We modeled vacancy luminescence with a double Gaussian.  The first Gaussian described the copper vacancy luminescence.  The second Gaussian described the $V_{\ce{O}}^{2+}$ luminescence.  There are two limitations of the model:  First, as in previous studies, silicon based detectors do not detect the low energy tail of the copper vacancy luminescence very efficiently.  Second, the $V_{\ce{O}}^{1+}$ luminescence appears indistinctly between the other two peaks.  The oxygen vacancy luminescence line shape has been described in more detail using a multiple phonon emission model \cite{koirala2014relaxation}.

\section{Results and Discussion}
\subsection{Exciton Luminescence}
\begin{figure}
	\begin{center}
	\includegraphics[width=.7\textwidth]{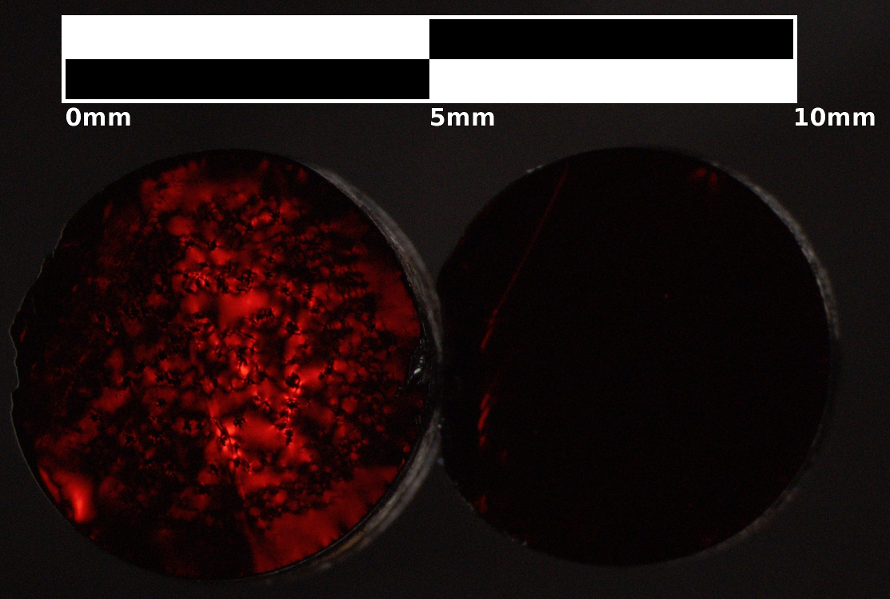}
	\end{center}
	\caption{Microscope image of back illuminated samples between perpendicular polarizers at room temperature.  The as-grown sample to the left exhibits birefringence (red) owing to strain.  The annealed sample to the right does not show birefringence because annealing removes strain.  Unpolarized images of the same samples can be found in \cite{chang2013removal}.}
	\label{fig:sample}
\end{figure}

The cuprous oxide lattice has octahedral $O_h$ symmetry (not chiral $O$ symmetry).  Crystals with this point group are not birefringent.  Fig. \ref{fig:sample} shows that the as-grown sample had birefringence near its cupric oxide inclusions, presumably owing to interfacial stress that lead to a local deformation of the crystal system \cite{haydar1980excitonic}, possibly to a tetragonal system.  The annealed sample did not show birefringence for two reasons: First, annealing removes cupric oxide inclusions \cite{chang2013removal} and second, annealing relieves stress.  These samples were selected for detailed study with the expectation that annealing would improve the quality of the exciton luminescence because strain would be removed.

For a strain tensor $\tilde{\varepsilon}$, the strain terms of the exciton Hamiltonian are \cite{sandfort2011paraexcitons,trebin1981excitons}
\begin{align}
	a \operatorname{Tr} \tilde{\varepsilon}+3b\varepsilon_{ii}\left( L_i^2-\frac13{\bf L}^2 \right)-\sqrt{3}d \varepsilon_{ij}\left( L_iL_j+L_jL_i \right),\label{stress}
\end{align}
where $a,b,d$ are the strain deformation potentials which are given for cuprous oxide in Table \ref{tab:def} and ${\bf L}$ is the angular momentum operator.  The strain perturbation leads to a splitting of the 1s orthoexciton polariton triplet energy level into three singlet energy levels.  The strain/splitting relationship can be used to estimate the size of the strain tensor elements from the energy splittings. Based on Table \ref{tab:def}, strain tensor elements are estimated to typically be less than $6\times 10^{-4}$.

\begin{table}
        \centering
        \begin{tabular}{rD{.}{.}{2}}
                Symbol &\multicolumn{1}{c}{Value (eV)}\\\hline
                $a$&-2.15 \\
                $b$&-0.43 \\
                $d$&0.36
        \end{tabular}
        \caption{Deformation Potentials of Cuprous Oxide \cite{sandfort2011paraexcitons}.}
        \label{tab:def}
\end{table}

\begin{figure}
	\begin{center}
	\includegraphics[width=.7\textwidth]{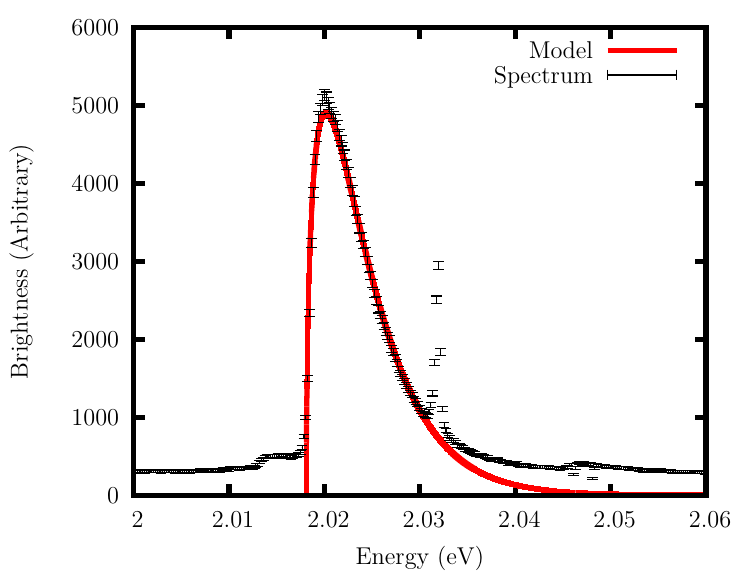}
	\end{center}
	\caption{An example of a typical luminescence spectrum from the annealed sample.  The model of the phonon-assisted $^2\Gamma_{12}^-$ luminescence is a Maxwell-Boltzmann distribution starting at 2.0181 eV with a temperature of 48 Kelvin (Eq. \ref{mb}). The narrow peak is exciton polariton luminescence.  The position on the sample is (0, 0) in Figure \ref{fig:energy}.}
	\label{fig:example}
\end{figure}

\begin{figure}
	\begin{center}
	\includegraphics[width=.7\textwidth]{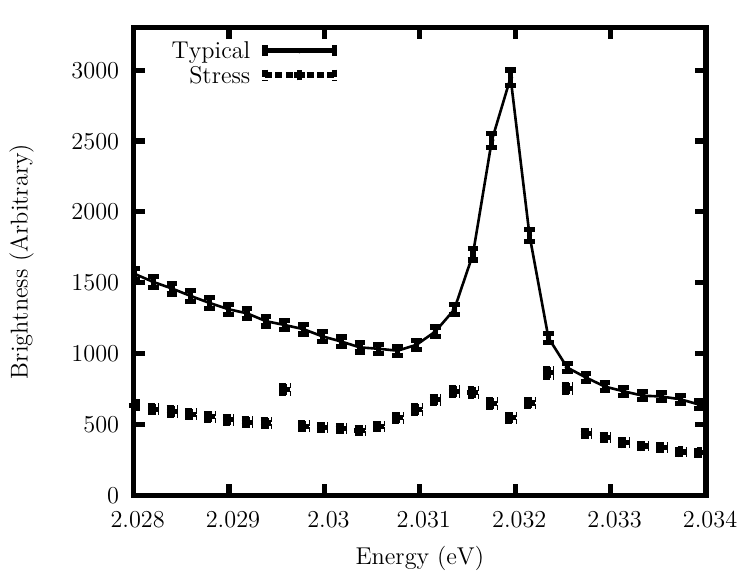}
	\end{center}
	\caption{A typical, resolution limited spectrum of polariton luminescence and an atypical spectrum showing three peaks due to stress.  Both are from the annealed sample.  The position of the three peaked outlier spectrum on the sample is (480, 460) $\mu$m in Figure \ref{fig:energy}.}
	\label{fig:polariton}
\end{figure}

Typical spectra, such as the one in Fig. \ref{fig:example}, did not show strain splitting in the polariton luminescence or the phonon-assisted luminescence.  However, we did find a location where the polariton luminescence was slightly split; this spectrum is the lower one in Fig. \ref{fig:polariton}.  The observed splitting pattern is the same as the pattern observed from a previous study where strain was applied to a sample \cite{sun2002strain}. 

\begin{figure}
	\begin{tabular}{cc}
	\includegraphics[width=.48\textwidth]{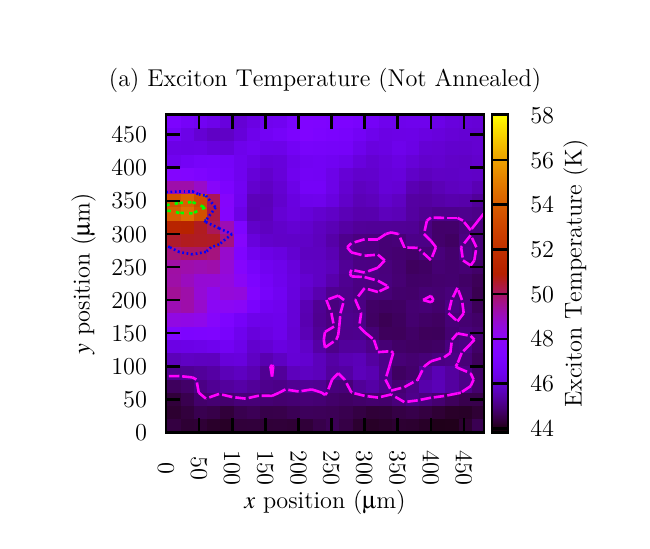}&\includegraphics[width=.48\textwidth]{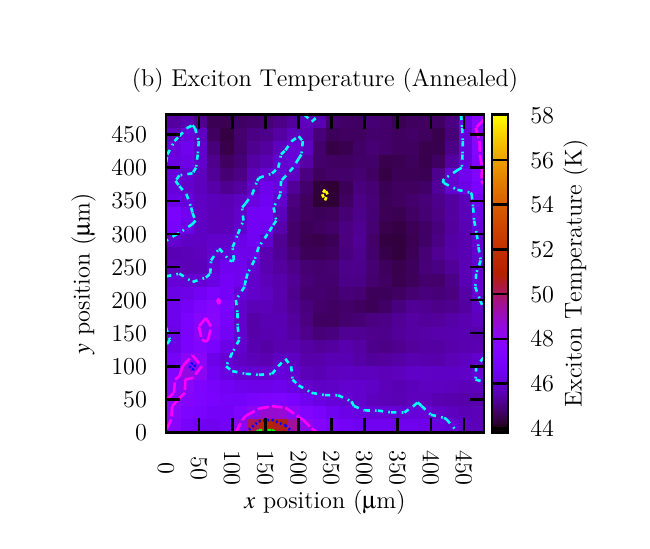}\\
	\includegraphics[width=.48\textwidth]{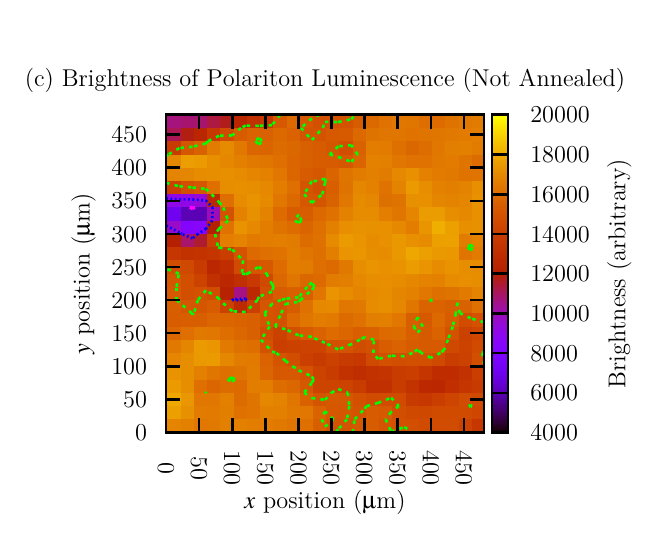}&\includegraphics[width=.48\textwidth]{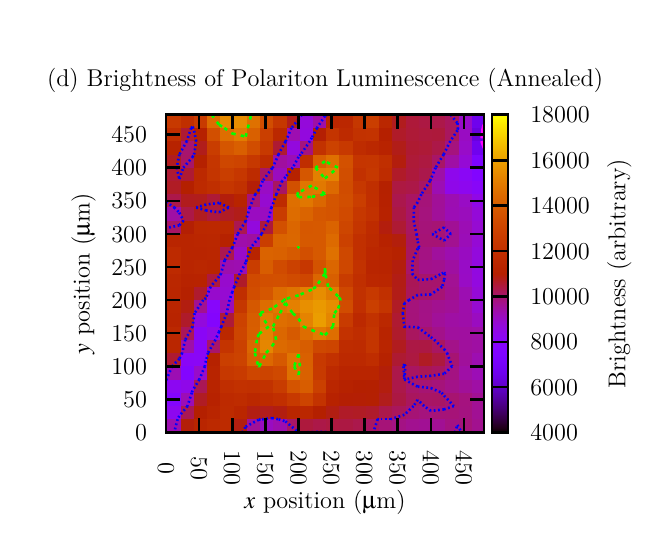}\\
	\includegraphics[width=.48\textwidth]{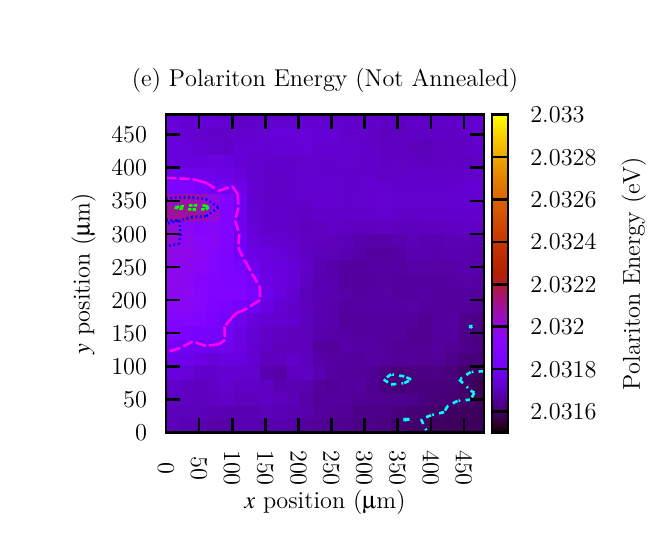}&\includegraphics[width=.48\textwidth]{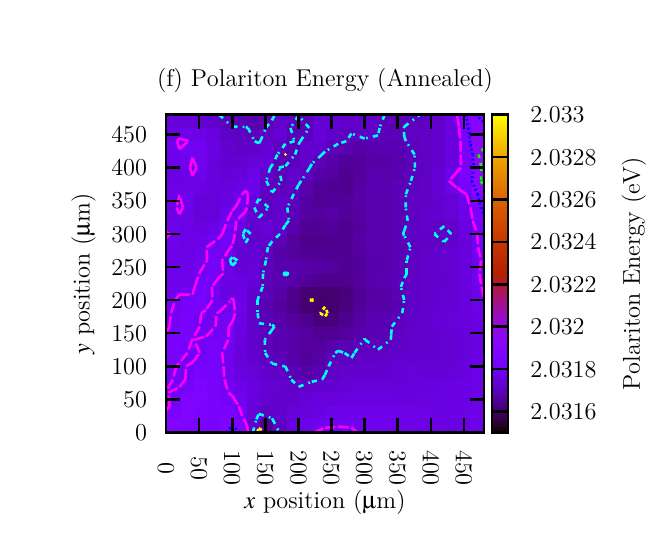}
	\end{tabular}
	\caption{
		(a, b) Spatially resolved temperature of the exciton gas.  Temperature is determined using the Maxwell-Boltzmann model (Eq. \ref{mb}) shown in Fig. \ref{fig:example} for the spectrum at each location.  The excitons are not expected to be in equilibrium with the lattice.  In each sample, a few defects mildly increase the exciton temperature measurement.
	(c, d) Brightness of the orthoexciton polariton luminescence.  Defects reduce the brightness of the luminescence.  Brightness is not comparable between samples.
	(e, f) Energy of the orthoexciton polariton luminescence. Stress defects can lead to tiny shifts in the mean luminescence energy.  (c, d) is compared with (e, f) to produce the pattern in Fig. \ref{fig:poltemp}.
	(a, c, e) Sample without annealing.  (b, d, f) Sample with annealing.
	}
	\label{fig:temp}
	\label{fig:amp}
	\label{fig:energy}
\end{figure}

The exciton temperature as determined from regression of Eq. \ref{mb} was typically about 46 K. Some locations show a higher temperature, which we attribute to defects.   Figs. \ref{fig:temp} (a, b) are maps of the exciton temperature.  The exciton temperature is higher than the thermometer temperature primarily because excitons do not completely thermalize with the crystal lattice before decaying.  In addition, there may be a temperature gradient across the sample since it is heated by the laser.

Defects in the locations with elevated temperatures also reduce the brightness of the orthoexciton polariton luminescence. Figs. \ref{fig:amp} (c, d) are maps of that brightness.  The brightness could be reduced because of lower transmission of the crystal surface, or it could be because of excitons decaying in some alternative way at defects.  Fig. \ref{fig:energy} (e, f) shows there are also very small shifts in the energy of the polariton luminescence from place to place which may be explained by stress in the sample (Eq. \ref{stress}).  Annealing reduced the variation in the mean polariton energy and in $E_c$ across positions (one-tailed F-tests \cite{press2007numerical}), consistent with our expectation that annealing would reduce strain in the sample.

\begin{figure}
	\includegraphics[width=.5\textwidth]{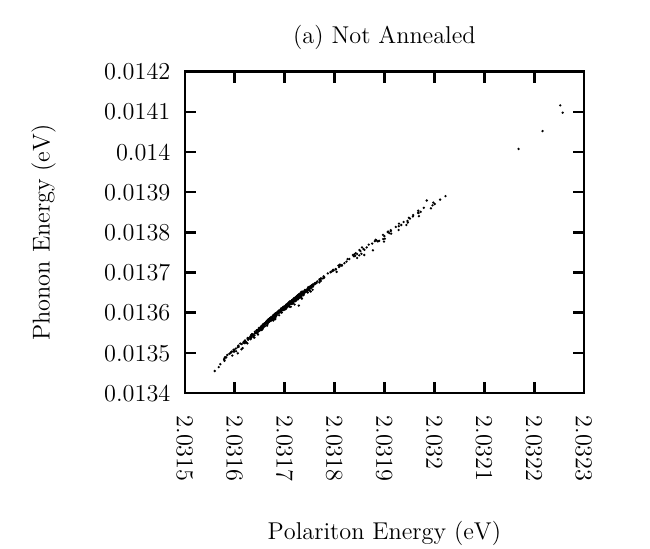}
	\includegraphics[width=.5\textwidth]{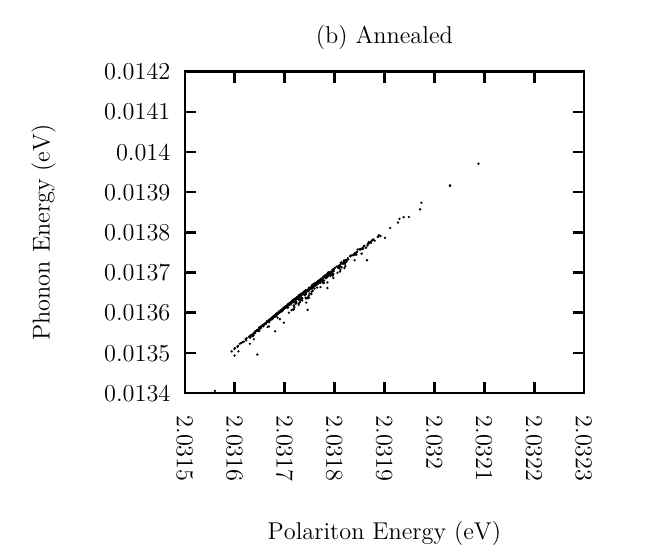}
	\caption{The phonon energy is inferred from the polariton energy minus the onset energy of phonon-assisted luminescence. Phonon energy positively correlates with polariton energy.  Annealing reduces the spread in the polariton luminescence energy  and the spread of the onset energy of phonon-assisted luminescence $E_c$ (one-tailed F-tests).  
	(a) Sample without annealing.  (b) Sample with annealing.
	}
	\label{fig:phononenergy}
\end{figure}

The phonon-assisted luminescence is caused by the decay of an exciton into a phonon and a photon. For a phonon with energy $E_\Gamma$, $E_c$ in Eq. \ref{mb} should be related to the polariton energy $E_p$ by the consistent energy equation 
\begin{align}
	E_p&=E_c+E_\Gamma+E_{KE}.
\end{align}
Based on a polariton wavenumber of $2.63\times 10^7/$m \cite{frohlich1991coherent}, the difference between the polariton energy and the exciton ground state energy $E_{KE}\approx 0.01$ meV.  Therefore we tentatively assume $E_{KE}$ is negligible.  
In Fig. \ref{fig:phononenergy}, the phonon energy is determined from $E_p-E_c$.  Surprisingly, the inferred phonon energy is strongly and positively correlated with the polariton energy.  There is a tiny downward curvature in the data.  The typical energy difference $E_p-E_c$ is 0.0136 eV.

\begin{figure}
	\includegraphics[width=.5\textwidth]{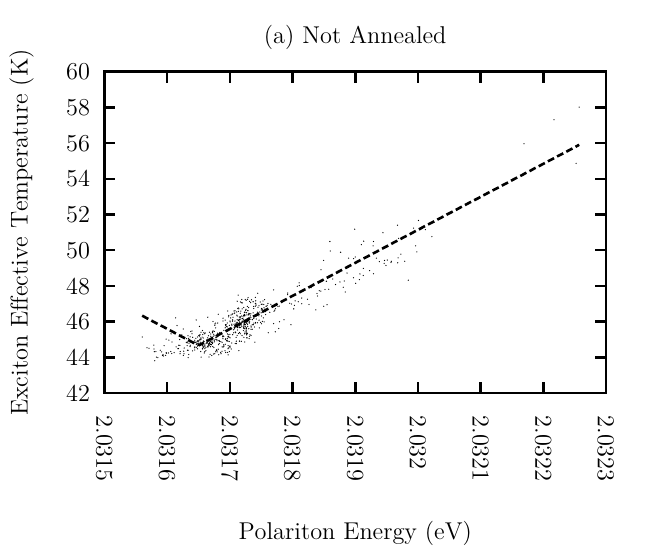}
	\includegraphics[width=.5\textwidth]{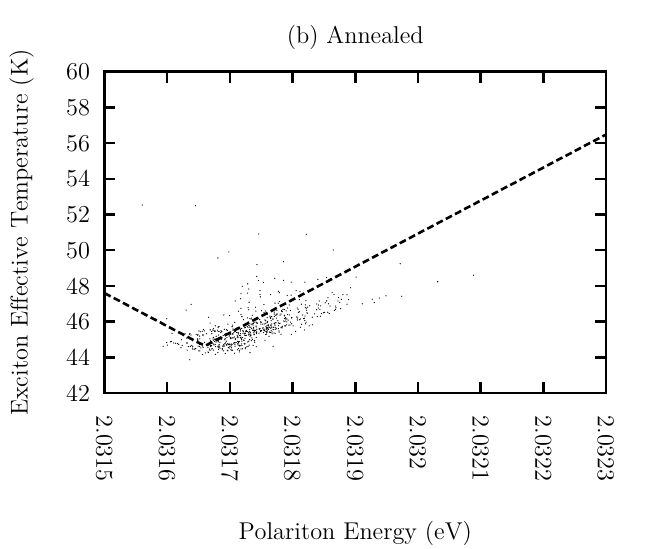}
	\caption{Strain can cause deviations from the intrinsic exciton polariton energy.  It can also cause a decrease in exciton lifetime.  Shorter exciton lifetimes imply less time for excitons to thermalize with the lattice, which leads to higher temperatures. The curve is based on Eq. \ref{thermal}.}
	\label{fig:poltemp}
\end{figure}

Strain increases the radiative transition rate, which reduces the exciton lifetime.  For example, the $D_{2h}$ group has more allowed quadrupole operators than the $O_h$ group \cite{liu2004excitons}.  If the intrinsic exciton polariton energy is $E_0$ and the intrinsic lifetime is $t_0$, but the measured energy is $E$, then a locally linear approximation of the expected decay rate for strains much less than the yield stress is 
\begin{align}
	\frac{1}{t}&=\frac{1}{t_0}+c|E_0-E|\\
	t(E)&=\frac{t_0}{{1}+ct_0|E_0-E|},
\end{align}
where $c$ is a constant.  If the exciton lifetime is infinite, the excitons thermalize to the lattice temperature $T_l$.  A value $T_i$ characterizes the initial condition of the excitons before they have thermalized.  If $\tau$ is the thermalization time constant, then for exciton subpopulations that exist for a duration $t_d$,
the temperature of the decaying subpopulation
according to the heat equation is
\begin{align}
	T(t_d)&=(T_i-T_l)e^{-t_d/\tau}+T_l\label{cool}
\end{align}
for $t_d$ much longer than the spontaneous phonon emission time.   
The temperature determined from decay luminescence can be approximated by an average of Eq. \ref{cool} over $t_d$ weighted by the probability density of exciton decay:
\begin{align}
	\braket{T(t(E))}&=\int_0^\infty T(t_d)
	\frac{e^{-\frac{t_d}{t(E)}}}{t(E)}dt_d
	\end{align}
	Integrating, substituting, and performing a Taylor expansion,
	\begin{align}
		\braket{T(E)}&=\left( T_i-T_l \right)\tau\left(\frac{1}{t_0+\tau}+\frac{ct_0^2|E_0-E|}{\left( t_0+\tau \right)^2}+\mathcal{O}\left( \left(ct_0|E_0-E|\right)^2 \right)  \right)+T_l\label{thermal}.
\end{align}
This relationship is not valid for short lifetimes, for large strains, or when excitons interact more strongly with each other than with the lattice.
Fig. \ref{fig:poltemp} shows the first two terms of expansion \ref{thermal} are sufficient to explain the data.  
$E_0$ was 2.0317 eV.  Most measurements showed polariton energies higher than this value.  

\begin{figure}
	\includegraphics[width=.5\textwidth]{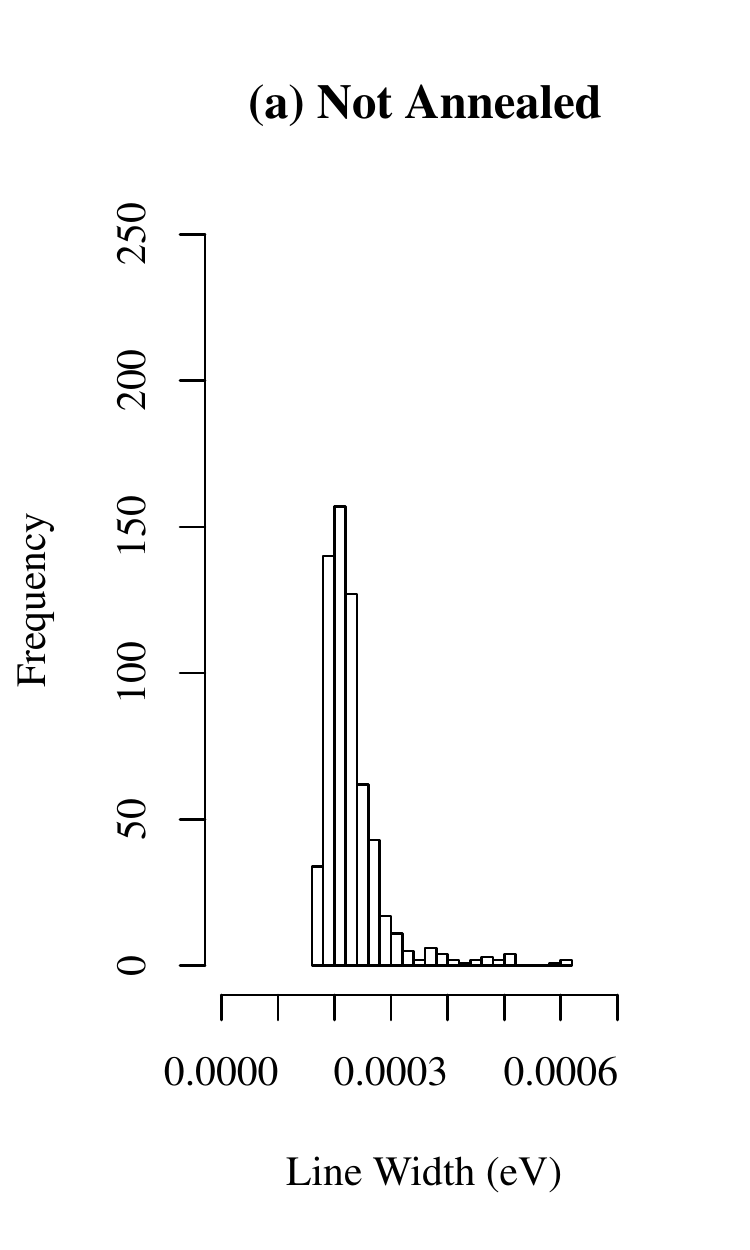}
	\includegraphics[width=.5\textwidth]{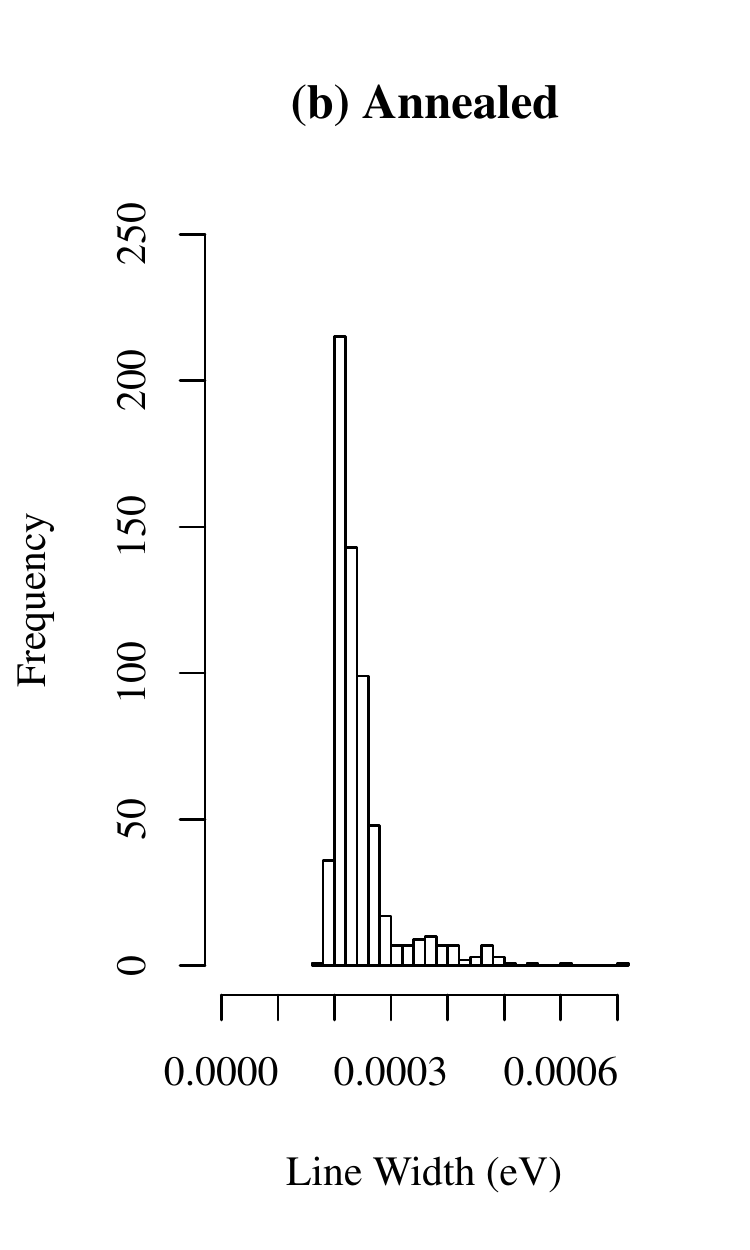}
	\caption{Histograms of polariton line width.  In most cases, the line width is consistent with the instrument resolution.  (a) Sample without annealing.  (b) Sample with annealing.}
	\label{fig:hist}
\end{figure}

Fig. \ref{fig:hist} is a histogram of the measured exciton polariton line widths, as defined by the residual brightness square weighted standard deviation energy. The polariton luminescence line width is typically consistent with the instrument resolution. However, there are a few outliers with a substantially larger measured line width.  A large line width may indicate a large local variation in strain.  

\subsection{1.95 eV Luminescence}

\begin{figure}
	\begin{center}
	\includegraphics[width=.7\textwidth]{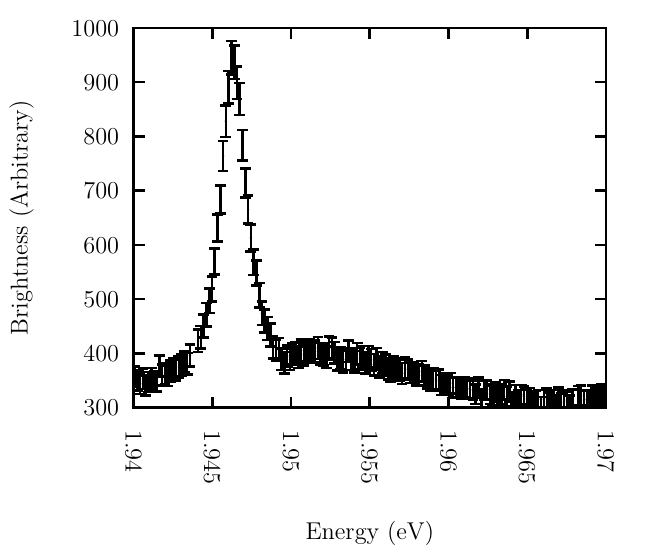}
	\end{center}
	\caption{An example of two phonon emission luminescence from the annealed sample, picked from a particularly bright location.  The main peak is the proposed two $^1\Gamma_2^-$ phonon emission luminescence.  The slight hump to the right is caused by $^3\Gamma_{15}^-$ phonon emission luminescence.  The position on the sample is (0, 40).}
	\label{fig:exdouble}
\end{figure}

\begin{figure}
	\includegraphics[width=.5\textwidth]{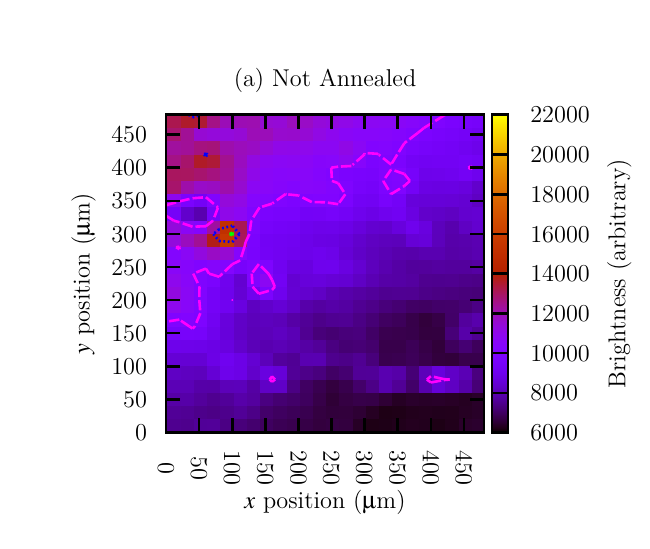}
	\includegraphics[width=.5\textwidth]{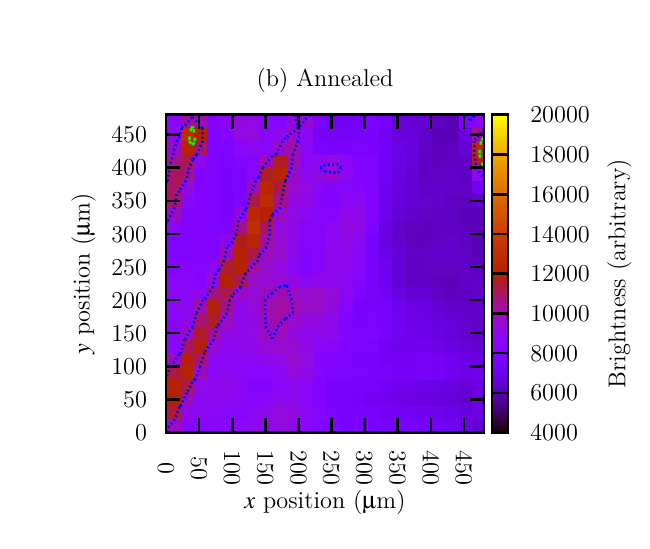}
	\caption{Intensity of two phonon emission luminescence at 1.95  
	eV showing bright defects.  (a) Sample without annealing.  (b) Sample with annealing.}
	\label{fig:double}
\end{figure}

The luminescence at 1.95 eV is typically very weak \cite{petroff1975study}.  Our spatially resolved spectra show that this is caused by spatial averaging.  The brightness can be larger at isolated  locations.  Fig. \ref{fig:exdouble} is the spectrum at a particularly bright spot. The map in Fig. \ref{fig:double} shows that the luminescence is localized. 

Petroff et al. \cite{petroff1975study} did not observe the luminescence above 4.2 K, which is slightly below the operating temperature for this experiment.  Petroff identifies three two-phonon emission processes and one three-phonon emission process, but suggests that a lack of temperature broadening indicates that the 1.95 eV luminescence is not caused by phonon emission.  

We agree with Petroff that the cause of the luminescence is extrinsic, but point out that energetically it is consistent with emission of two 0.0431 eV  $^1\Gamma_2^-$ phonons \cite{petroff1975study}.  This two phonon emission luminescence's energy lies just below the peak of the weak, broad $^3\Gamma_{15}^-$ phonon emission luminescence, which can also be found in Fig. \ref{fig:exdouble}.  Two-phonon assisted luminescence is present in the decay of cuprous oxide paraexcitons in high magnetic fields \cite{sandfort2008resonant}.  Two phonon transitions are also known in silicon \cite{dumke1960two} and graphene \cite{basko2009electron}.  An additional extrinsic feature has been reported at about 1.91 eV in other work \cite{prevot1972near,zouaghi1972near2,zouaghi1972near,gastev1982relaxed,petroff1972luminescence}.


\begin{figure}
	\includegraphics[width=.5\textwidth]{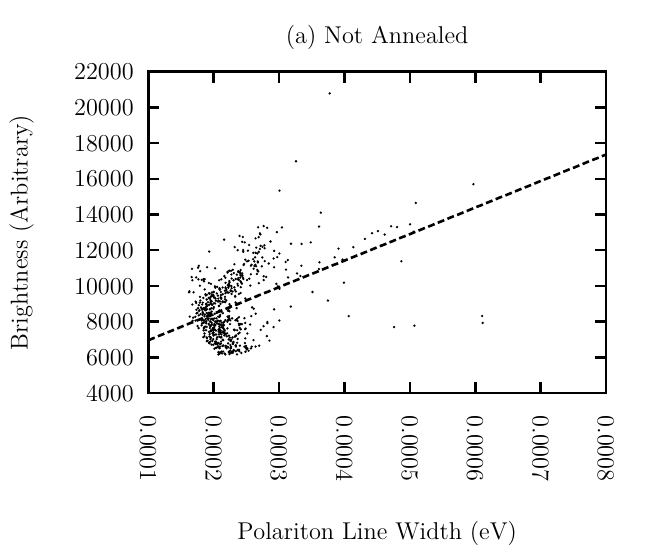}
	\includegraphics[width=.5\textwidth]{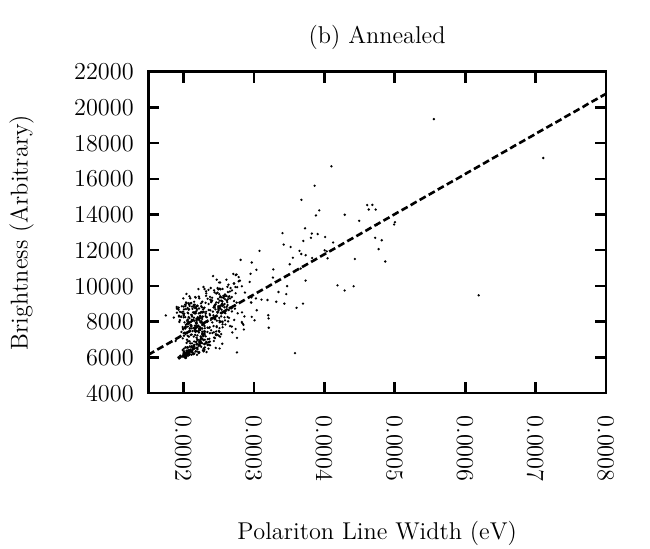}
	\caption{Brightness of two phonon emission luminescence versus polariton line width.  The brightness increases as strain breaks the parity symmetry of the exciton.  The polariton line width also increases for the same reason.  (a) Sample without annealing.  (b) Sample with annealing.
	}
	\label{fig:doublewidth}
\end{figure}

Hypothesizing that this localized luminescence may be caused by strains in the sample or by defects which cause strain, we expect to find a correlation between the spectral width of the exciton polariton and the brightness of the 1.95 eV two phonon emission luminescence.  As shown in Fig. \ref{fig:doublewidth}, a significant positive correlation was found, however the coefficient of determination was only $R^2=0.2$ for the as-grown sample and $R^2=0.6$ for the annealed sample.  Since the exciton polariton line width is normally less than the instrument resolution, it is challenging to find strong relationships.

%\begin{figure}
%	\includegraphics[width=.5\textwidth]{<++>}
%	\includegraphics[width=.5\textwidth]{<++>}
%	\caption{<++>.}
%	\label{fig:<++>}
%\end{figure}

\subsection{Vacancy Luminescence}

\begin{figure}
	\begin{center}
	\includegraphics[width=.7\textwidth]{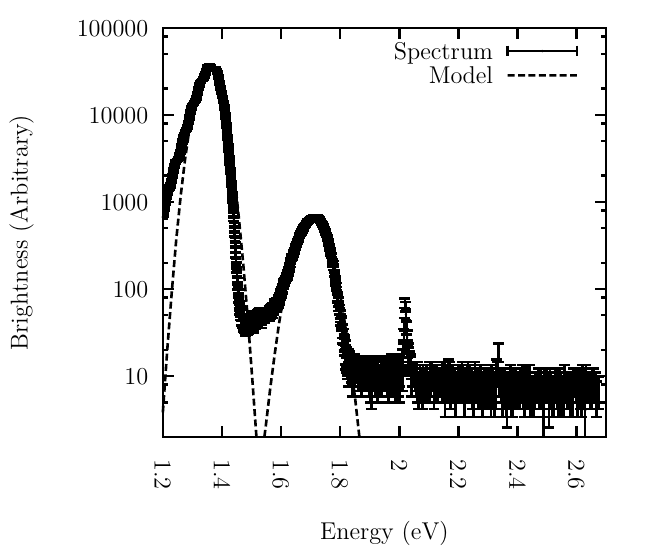}
	\end{center}
	\caption{An example of a typical copper and oxygen vacancy luminescence spectrum from the annealed sample modeled with two Gaussians. The position on the sample is (0, 0).}
	\label{fig:exvac}
\end{figure}

Our samples primarily show copper vacancy luminescence (1.4 eV), consistent with other studies using crystals grown with the floating zone method.  Higher energy oxygen vacancy luminescence (1.7 eV) is also present.  Fig. \ref{fig:exvac} shows that the copper vacancy luminescence is over an order of magnitude stronger.  We only observe the doubly ionized oxygen vacancy luminescence clearly.  The singly ionized oxygen vacancy luminescence is mostly obscured by the two overlapping, stronger peaks.  In the coarse detection mode, the exciton polariton and phonon-assisted luminescence are also apparent, but they are not well resolved.  Leakage of laser light through the long wavelength pass filter is also barely detected.

\begin{figure}
	\includegraphics[width=.5\textwidth]{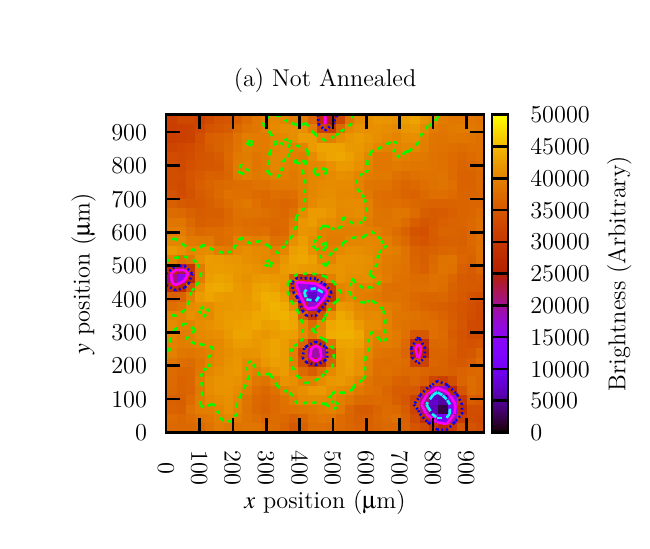}
	\includegraphics[width=.5\textwidth]{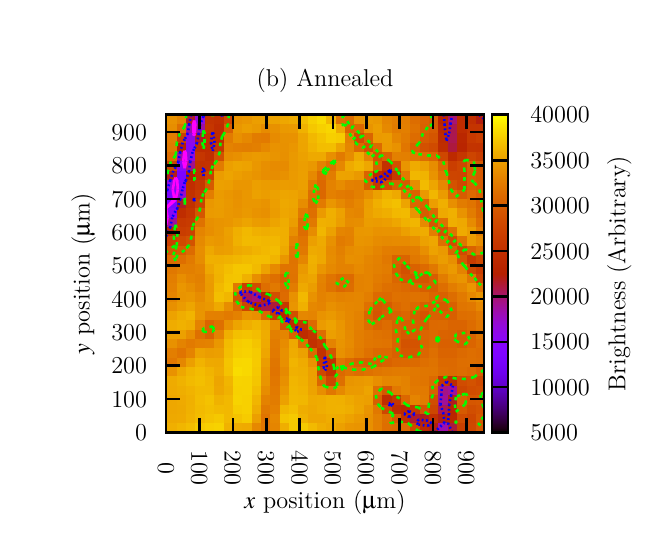}
	\caption{Spatially resolved brightness of copper vacancy luminescence.  (a) Sample without annealing.  (b) Sample with annealing.}
	\label{fig:copper}
\end{figure}

\begin{figure}
	\includegraphics[width=.5\textwidth]{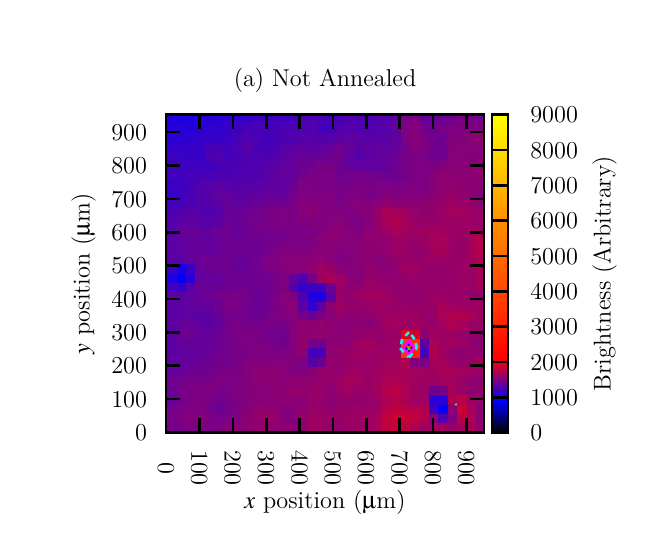}
	\includegraphics[width=.5\textwidth]{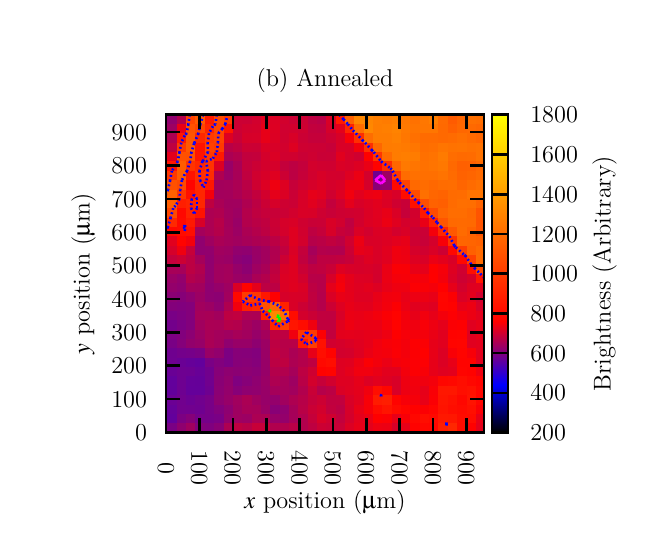}
	\caption{Spatially resolved brightness of oxygen vacancy luminescence.  Compare (b) with Fig. \ref{fig:copper} (b) to see the negative correlation in Fig. \ref{fig:cuovac} (b). (a) Sample without annealing.  (b) Sample with annealing.}
	\label{fig:oxygen}
\end{figure}

The map in Fig. \ref{fig:copper} shows some locations on the samples have decreased copper vacancy luminescence.   Copper vacancies may be removed near cupric oxide inclusions through the reaction \cite{chang2013removal}
\begin{align}
	\cee{$V_{\ce{Cu}}^{\cco}$ + $\ce{Cu}_{\ce{Cu}}^{\cco}$ +$\ce{O}_{\ce{O}}^{\cco}$&-> $\ce{Cu}_{\ce{Cu}}^{\ce{CuO}}$ + $\ce{O}_{\ce{O}}^{\ce{CuO}}$}.\label{removecuvacancies}
\end{align}
Fig. \ref{fig:oxygen} shows maps of the oxygen vacancy luminescence brightness.  Oxygen vacancies  coexist with copper vacancies because cuprous oxide forms under nonequilibrium conditions, where the degree of net copper deficiency varies from place to place.  

\begin{figure}
	\includegraphics[width=.5\textwidth]{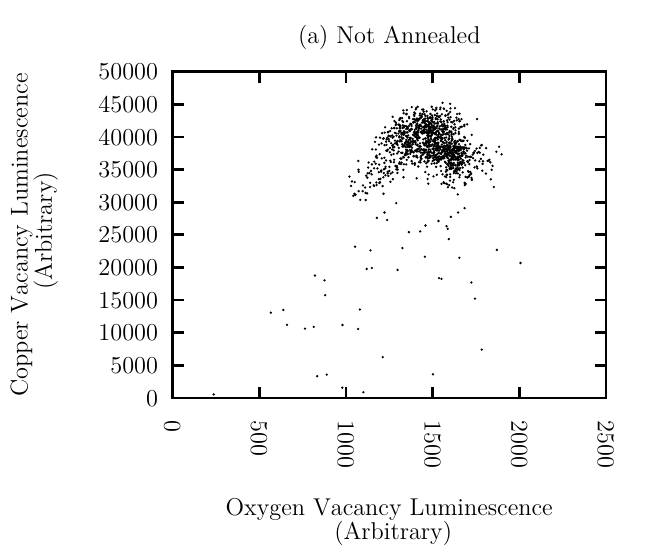}
	\includegraphics[width=.5\textwidth]{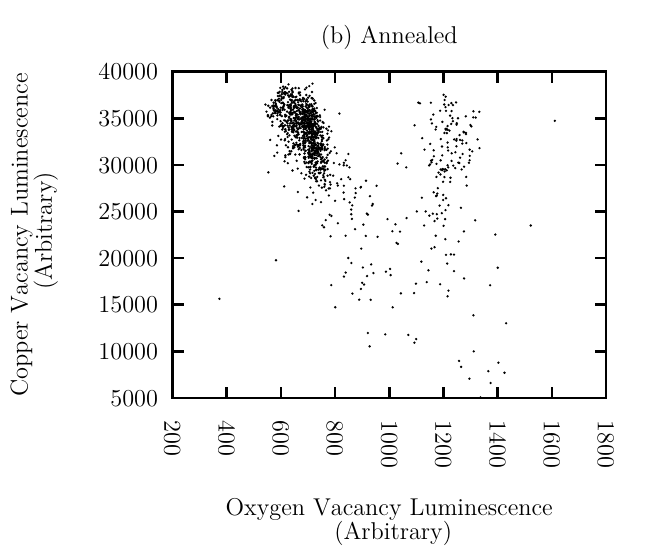}
	\caption{Comparing the brightness of luminescence from copper and oxygen vacancies.  After annealing, a negative correlation develops as the two types of vacancies annihilate.  (a) Sample without annealing.  (b) Sample with annealing.  Two data points are not shown in (a); their values are (8620, 34819) and (3106, 37012).}
	\label{fig:cuovac}
\end{figure}

Fig. \ref{fig:cuovac} shows there is a negative correlation across locations between copper and oxygen vacancies in the annealed sample with $R^2=0.3$, but a weak positive correlation in the as-grown sample with $R^2=0.1$.  There are four ways stoichiometry in cuprous oxide can potentially be manipulated.  The first two are interfacial exchange between \cco and  \ce{Cu} or \ce{O}.  The third way, exchange with \ce{CuO}, is facilitated by copper vacancies (Reaction \ref{removecuvacancies}).  The fourth possibility is annihilation of unlike vacancies:
\begin{align}
	\cee{$V_{\ce{O}}^{\cco}$ +2 $V_{\ce{Cu}}^{\cco}$& -> \text{null}}.\label{annihilation}
\end{align}
The as-grown sample was cooled rapidly to room temperature after crystallizing, preventing much diffusion from occurring.  
When the annealed sample was cooling, the vacancies in the lattice diffused.  Annihilation is consistent with the development of a negative correlation after annealing.  The remaining vacancy concentrations are determined by the spontaneous spatial variation in \ce{Cu} deficiency, but the annealing and annihilation increases the length scale and magnitude of deficiency variations relative to overall deficiency.


\section{Conclusions}

Defect luminescence is a useful tool for evaluating advances in cuprous oxide synthesis \cite{ohkubo2014nanospace,wu2014octahedral,labidi2014synthesis,li2013engineering,sun2014recent}.  Detailed luminescence imaging of cuprous oxide crystals grown by the floating zone method shows that there are few defects in the samples.  Sample stress plays an important role in exciton dynamics and luminescence.  Based on Table \ref{tab:def}, strain tensor elements are estimated to typically be less than $6\times 10^{-4}$.  Our as-grown sample's exciton polariton luminescence showed little evidence of stress, but the annealed sample was even better.  The 1.95 eV luminescence appears to be associated with defects in the crystal and is localized in bright spots.  Annealing cuprous oxide crystals contributes to the annihilation of copper and oxygen vacancies, leading to a negative correlation between the two kinds of vacancy associated luminescence.  
\section{Acknowledgments}
L. F., E. L., and J. K. gratefully acknowledge NSF IGERT DGE-0801685 and funding by the Institute for Sustainability and Energy at Northwestern (ISEN).
Crystal growth was supported by NSF DMR-1307698 and in part by Argonne National Laboratory under U.S. Department of Energy contract DE-AC02-06CH11357.
K. C. was supported as part of the Center for Inverse Design, an Energy Frontier Research Center funded by the U.S. Department of Energy, Office of Science, Office of Basic Energy Sciences, under award number DE-AC36-08GO28308.
This work made use of the OMM and Magnet, Low Temperature, and Optical Facilities supported by the MRSEC program of the NSF (DMR-1121262) at the Materials Research Center of Northwestern.
N. P. S. acknowledges support as an Alfred P. Sloan Research Fellow.
\section*{References}

\bibliography{j}

\end{document}